\documentstyle[epsfig,12pt]{article}

\newcommand {\eqref} [1] {(\ref {#1})}
\newcommand {\beq} {\begin{equation}} 
\newcommand {\eeq} {\end{equation}}
 \newcommand {\ber}{\begin{eqnarray*}}
 \newcommand {\eer} {\end{eqnarray*}}
\newcommand {\bea}{\begin{eqnarray}}
 \newcommand {\eea} {\end{eqnarray}}

\newcommand{\drawsquare}[2]{\hbox{%
\rule{#2pt}{#1pt}\hskip-#2pt
\rule{#1pt}{#2pt}\hskip-#1pt
\rule[#1pt]{#1pt}{#2pt}}\rule[#1pt]{#2pt}{#2pt}\hskip-#2pt
\rule{#2pt}{#1pt}}

\newcommand{\fund}{\raisebox{-.5pt}{\drawsquare{6.5}{0.4}}}
\newcommand{\antifund}{\overline{\fund}}

\def\ov{\overline} 

\def\Acknowledgements{\bigskip  \bigskip {\begin{center} \begin{large}
             \bf ACKNOWLEDGEMENTS \end{large}\end{center}}}

\begin{document}\begin{titlepage}
\rightline{CERN-TH/2002-386}
\rightline{IFT UAM-CSIC/02-58}
\rightline{FT-UAM-02/33}
\rightline{\tt hep-th/0301099}
\vskip .5cm
\centerline{{\Large \bf Closed Strings Tachyons and }} 
\vskip 0.3cm
\centerline{{\Large \bf Non-Commutative Instabilities}}
\vskip 1cm
\centerline{Adi Armoni\ ${}^\dagger$, Esperanza L\'opez\ ${}^\ddagger$
and Angel M. Uranga\ ${}^\diamond$}
\vskip 0.5cm
\centerline{${}^\dagger$ Theory Division, CERN}
\centerline{CH-1211 Geneva 23, Switzerland}
\centerline{adi.armoni@cern.ch}
\vskip 0.3cm
\centerline{${}^\ddagger$ Departamento de F\'{\i}sica Te\'orica C-XI
and Instituto de F\'{\i}sica Te\'orica  C-XVI}
\centerline{Universidad Aut\'onoma de Madrid, Cantoblanco, 28049 Madrid, Spain}
\centerline{Esperanza.Lopez@uam.es}
\vskip 0.3cm
\centerline{${}^\diamond$ IMAFF and Instituto de F\'{\i}sica Te\'orica  C-XVI}
\centerline{Universidad Aut\'onoma de Madrid, Cantoblanco, 28049 Madrid, Spain}
\centerline{Angel.Uranga@uam.es}

\vskip .5cm

\begin{abstract}
We observe a relation between closed strings tachyons and one-loop 
instabilities in non-supersymmetric non-commutative gauge theories.
In particular we analyze the spectra of type IIB string theory on 
$C^3/Z_N$ orbifold singularities and the non-commutative field theory that 
lives on D3 branes located at the singularity. We find a surprising 
correspondence between the existence or not of one-loop low-momentum 
instabilities in the non-commutative field theory and the existence or 
not of tachyons in the closed string twisted sectors. Moreover, the 
relevant piece of the non-commutative field theory effective action is
suggestive of an exchange of closed string modes. This suggests that 
non-commutative field theories retain some information about the dynamics 
of the underlying string configuration. Finally, we also comment on a 
possible relation between closed string tachyon condensation and field 
theory tachyon condensation.
\end{abstract}
\end{titlepage}

\section{Introduction}
\noindent

The issue of closed string tachyon condensation is one of the most
difficult problems in string theory. Bosonic string theory contains a
tachyon, but it does not necessarily mean that the theory is
inconsistent, although the ultimate fate of its decay process is still 
mysterious.

Recently there has been a lot of progress in the understanding of tachyon
condensation processes in non-supersymmetric string theories with
worldsheet supersymmetry \cite{Adams:2001sv,David:2001vm}. The prime 
example is IIB string theory on geometric orbifold singularities that 
break spacetime supersymmetry. In this case the twisted sector tachyons 
live on the singularity. In several cases, it was argued that the twisted 
sector tachyons condense and the true vacuum of the theory is type IIB on 
flat background \cite{Adams:2001sv}.
Another example is type 0 string theory, which can be thought of as an 
orbifold of type II. The tachyon in this case is ten-dimensional.
Although the meaning of this instability is not completely clear, there 
are arguments in favor of this tachyon condensation process having type II 
theory as the corresponding endpoint \cite{David:2001vm}.

Another, seemingly unrelated subject, is the study of non-commutative  
gauge field theories (see \cite{Douglas:2001ba,Szabo:2001kg} for recent 
reviews on the subject). These theories exhibit many similarities to open 
string theory: they contain dipoles as fundamental objects, there is a 
connection between the UV and the IR (UV/IR mixing)
\cite{Minwalla:2000px,Matusis:2000jf} and the gauge group and its 
representation are constrained
\cite{Terashima:2000xq,Matsubara:2000gr,Armoni:2000xr,Hayakawa:1999yt}. In addition, due to 
lack of Lorentz invariance the gauge field often becomes tachyonic (or 
massive), since both gauge invariance and Lorentz symmetry are needed in 
order to protect the masslessness of the gauge boson
\cite{Ruiz:2001hu,Landsteiner:2001ky,Guralnik:2002ru}. More precisely, 
one-loop effects modify the dispersion relation of gauge bosons in the 
center of the group. The modes polarized along the non-commutative 
directions verify
\beq
E^2={\vec p}^2-\lambda {N_b-N_f \over \pi^2} {1 \over {\tilde p}^2} \, ,
\label{disp}
\eeq
where $\lambda$ is the 't Hooft coupling; 
${\tilde p}^\mu =\theta^{\mu \nu} p_\nu$ and $\theta^{\mu \nu}$ 
is a matrix that parametrizes the noncommutativity of the space 
$[x^\mu,x^\nu]_* = i\theta^{\mu \nu}$. To avoid problems with
unitarity \cite{Gomis:2000zz}, we will always consider
$\theta^{\mu \nu}$ with only space indices, i.e.
$\theta^{0i}=0$. In the previous equation $N_b$ and $N_f$ are the number 
of bosonic and fermionic degrees of freedom in the adjoint representation. 
In particular, when there are more bosons than fermions, the low
momentum modes of the gauge bosons in the center of the group become 
unstable. 

\medskip

In this work we consider non-commutative field theories arising on 
the world-volume of D-branes in the decoupling limit, and observe a 
relation between the non-commutative field theory instabilities and 
the closed string tachyons in the parent string theory. In a previous 
work \cite{Armoni:2001uw} it was found that the piece of the one-loop 
effective action containing the instability, for the non-commutative 
field theory arising on a collection of D3 branes of type 0 string 
theory, has a form suggestive of an origin from exchange of the closed 
string modes. There were also some hints that the instability associated 
to the gauge boson in the field theory is related to the presence of the 
bulk tachyon. However, the relation between the two could not be made 
precise.

In the present work we generalize this observation to an 
infinite class of models, and suggest that the two (seemingly very 
different!) kinds of instabilities are related. We consider type IIB
string theory on geometric orbifold singularities, with a constant 
background of NS-NS 2-form field in the untwisted directions.
We analyze two quantities: the spectrum of the lowest modes of the closed 
string tower in the different twisted sectors, and the one-loop correction 
to the dispersion relation of the gauge bosons in the non-commutative 
field theory that lives on a 
collection of $M$ D3 branes sitting at the orbifold singularity. 
The closed string contains $N$ sectors: untwisted sector and $N-1$ twisted 
sectors. The various low-lying modes (in the NS-NS sector) in each sector 
are either massless, massive or tachyonic. 
The generic non-commutative field theory is a $U(n)^N$ gauge theory. 
We can take linear combinations of the $N$ center gauge bosons
which naturally couple to the different closed string 
twisted sectors of the underlying string theory picture. These linear 
combinations are not mixed by the one-loop mass correction (they 
diagonalize the corresponding mixing matrix) and can remain massless at 
one-loop, or become massive or tachyonic - according to \eqref{disp}. 

We find that there is a one-to-one correspondence between the two systems, 
in the following sense: The non-commutative field theory produces one-loop 
instabilities precisely for those linear combinations coupling to closed 
string twisted sectors containing tachyons; the one-loop correction exists 
but does not lead to instabilities for linear combinations coupling to closed 
string twisted sectors which are non-supersymmetric but do not contain 
tachyons; finally, the one-loop correction is absent precisely for linear 
combinations coupling to closed string twisted sectors whose associated 
twist preserves some supersymmetry. 
Thus, the one-loop non-commutative dynamics that leads to this rich 
pattern of field theory mass corrections seems to correlate with 
properties of the closed sector of the underlying string theory. 
In other words, the non-commutative field theory seems to retain some 
crucial aspects of the closed string dynamics!

We moreover show that the full form of the one-loop effective action of 
the non-commutative theory is very reminiscent of the exchange of closed 
strings. This and the previous observation might lead to the speculation 
that tachyon condensation processes in the two systems are also related.
Unfortunately present techniques do not allow to explore the instability 
of non-commutative field theory to the point of really testing this 
beautiful, but far stronger, form of a correspondence

The organization of the paper is as follows: in section 2 we present our
main result. We analyze the (one-loop) field theory spectrum and we compare 
it to the (tree-level) lowest modes of the closed string spectrum. We show 
the correlation between the existence of non-commutative field theory 
instabilities and closed string twisted tachyons. In section 3 we show 
that the one-loop effective action for the field theory that lives on 
orbifold singularities is reminiscent of an exchange of twisted sector 
modes. Finally, in section 4 we discuss our results and comment about the
possibility of tachyon condensation in the field theory and the relation 
to closed strings dynamics.

\section{Closed Strings Spectrum versus Field Theory Spectrum}

\noindent

In this section we will compare the closed string spectrum of type IIB 
string theory on orbifold singularities to the field theory spectrum
on D3-branes sitting at the tip of the singularity. 
We will calculate first the one-loop correction to the mass of 
the gauge bosons in the field theory side and then we will compare it to 
the masses of the lowest modes in the closed string tower.

We consider type IIB theory on a generic $C^3 / Z_N$ orbifold (although 
we expect our results to extend straightforwardly to other orbifold 
groups, including non-abelian ones). Regarding 10d flat space as 
$M_4\times R^6$, the Lorentz group decomposes as $SO(1,9)\rightarrow 
SO(1,3) \times SO(6)$. The orbifold action is a 
discrete subgroup of this $SO(6)$. Its action can be specified by defining 
how it acts on the spinor representation $4$ of $SO(6)\sim SU(4)$. This 
is given by an order $N$ matrix in $SU(4)$, which can be taken 
diagonal
\begin{equation}
{\rm diag}(e^{2 \pi i a_1 \over N},e^{2 \pi i a_2 \over N},
e^{2 \pi i a_3 \over N},e^{2 \pi i a_4 \over N}) \, , 
\,\,\,\,\, \sum_\alpha a_\alpha = 0\,\, {\rm mod}\,\, N \, .
\end{equation}
The integers $a_\alpha$ are defined mod $N$.
Coordinates in $R^6$ transform in the vector representation $6$ of 
$SO(6)$, which is obtained from the antisymmetric tensor product of two 
$4$'s. Taking complex coordinates, the geometric action of $Z_N$ on the 
coordinates of $C^3$ is
\begin{equation}
Z^\beta \rightarrow e^{2 \pi i b_\beta \over N} Z^\beta \, \,\,\,\,\, 
\beta=1,2,3
\end{equation}
where the $b_\beta$ are determined by the $a_\alpha$ as
\begin{equation}
b_1=a_2 + a_3 \,\, , \,\,\,\,\, b_2=a_3 + a_1 \,\, , \,\,\,\,\,
b_3=a_1 + a_2 \, .
\label{ba}
\end{equation}
$C/Z_N$ and $C^2/Z_N$ orbifold models are obtained when two or one 
$b_\beta$'s respectively are zero. The case $b_\beta=0 ({\rm mod} N)$ but 
$a_\alpha$ non-trivial corresponds to type 0 string theory. Type 0 
orbifolds are also included among our models.

\medskip

We place now $n$ D3-branes \footnote{Here we refer to dynamical 
D3-branes, namely in the covering space, we consider $Nn$ D3-branes in 
$n$ copies of the regular representation.} at the fixed point of the 
orbifold. The 
tree-level spectrum of the field theory  on the branes is obtained from 
the massless states of the open string sector. Using standard projection 
rules for the Chan-Paton wavefunctions \cite{Douglas:1996sw}, the field 
theory contains the following fields. The gauge group is $U(n)^N$. The 
matter content consists of Weyl fermions transforming in the representation 
$(\fund_i,\antifund_{i+a_\alpha})$, for $\alpha=1,..,4$, and complex 
scalars in $(\fund_i,\antifund_{i+b_\beta})$, for $\beta=1,2,3$.
The theory is generically non-supersymmetric. Interactions are easily 
determined by keeping the $Z_N$-invariant terms in the action of the 
parent ${\cal N}=4$ supersymmetric theory of D3-branes in flat space. 
We will not need the explicit expressions here. However, it is important 
to realize that the gauge coupling constants of the different factors in 
the orbifold theory are all equal, since they are inherited from the 
unique gauge coupling in the parent theory of D3-branes in flat space. 
The same conclusion is reached by realizing that gauge couplings in the 
quotient theory are controlled by vevs of closed string twisted states, 
which are set to be equal at the point in moduli space which is described 
by a CFT orbifold. Being equal for all gauge factors, the gauge coupling 
$g$ will factor out of all our gauge field theory computations below.

Turning on a 2-form background in two (spatial) directions of the $M_4$ 
makes the corresponding field theory non-commutative. 
The models we consider here are generically chiral. In the commutative
case the $U(1)$'s are anomalous and therefore become massive via a
Green-Schwarz mechanism and
therefore decouple, leading to an $SU(n)^N$ theory at low-energies \cite{Ibanez:1998qp}. 
In the present non-commutative case the situation is more subtle
\cite{Armoni:2002fh}: the $U(1)$'s become massive and decouple only for zero
non-commutative momentum $\tilde p=0$. For non-zero $\tilde p$ the
anomaly vanishes. In the following we will consider only the latter
case.

One of the
most striking characteristics of non-commutative theories is the
appearance of infrared divergences whose origin is the integration of
high momenta in non-planar loops \cite{Minwalla:2000px}. We would like to 
study the effects of pole-like infrared divergences associated to a 
non-commutative theory with the previous gauge group and matter content. 
These poles generate a contribution to the polarization tensor of the 
$U(1)_i$ photons of the form $\Pi^{\mu \nu}_{ij}=M_{ij}
{g^2 n \over \pi^2} {{\tilde p}^\mu {\tilde p}^\nu \over \tilde p^4}$, 
with $i,j=1,..,N$. The matrix $M_{ij}$ can be read directly
from the matter content as follows. Only matter transforming in the
adjoint representation contributes to $M_{ii}$. In particular each 
bosonic degree of freedom contributes $+1$ and each fermionic degrees 
of freedom $-1$. When there is only one gauge group factor,
for example if only one type of fractional brane is placed at the 
orbifold point, this reproduces the 1-loop corrected dispersion
relation (\ref{disp}). $M_{ij}$ with $i \neq j$ is determined by 
the bifundamental matter content. The rule is as before: each
bosonic (fermionic) degree of freedom contributes $+1$ ($-1$).  
To our knowledge, the fact that bifundamental matter can give rise
to non-planar diagrams has not been mentioned in the existing literature.
However it is easy to see that this is the case if one uses 't Hooft's 
double line notation. It is possible to draw a non-planar graph for the 
vacuum polarization with bi-fundamental matter inside the loop, if the 
external legs belong to different $U(1)$ factors. The contribution to 
the 1-loop effective action that summarizes these effects is 
\begin{equation}
\Delta S ={g^2 \over 2 \pi^2} \int {d^4 p \over (2 \pi)^4}
\, {{\tilde p}^\mu {\tilde p}^\nu \over
{\tilde p}^4 } \sum_{i,j=1}^N M_{ij} \, {\rm Tr}\, A_\mu^{(i)}(p) 
{\rm Tr}\, A_\nu^{(j)}(-p) \, ,
\label{pole}
\end{equation}
where we have chosen the normalization for Lie algebra generators
${\rm Tr} t_a t_b=\delta_{ab}$, with $t_0={1 \over \sqrt{n}} {\bf 1}$.  
The matrix $M_{ij}$, as explained, is given by
\begin{equation}
M_{ij}= 2 \delta_{ij} - \sum_\alpha ( \delta_{i,j-a_\alpha}
+ \delta_{i,j+a_\alpha} ) + \sum_\beta ( \delta_{i,j-b_\beta}
+ \delta_{i,j+b_\beta} )  \, .
\label{polematrix}
\end{equation}
Note that in the above expression the vectors contribute twice as much as 
scalars, since they have two physical polarizations in four dimensions.
 
In order to analyze the potential instabilities arising from the 
pole-like contributions to the polarization tensor, we need to diagonalize 
the matrix $M$. This is achieved by the set of vectors $e^{(k)}$, 
$k=0,..,N-1$, with components $e^{(k)}_j=e^{2 \pi i \, {jk \over N}}$, 
$j=1,..,N$. The associated eigenvalues are
\begin{equation}
\epsilon^{(k)}=2\left(1-\sum_\alpha \cos {2 \pi k a_\alpha \over N}
+\sum_\beta \cos {2 \pi k b_\beta \over N}\right)=16 \prod_\alpha
\sin {\pi k a_\alpha \over N} \, ,
\label{eigenvalue}
\end{equation}
where, using the freedom to redefine the $a_\alpha$ modulo $N$, we have 
set $a_4=-(a_1+a_2+a_3)$. Thus the effective action can be rewritten as
\begin{equation}
\Delta S ={g^2 \over 2 \pi^2} \int {d^4 p \over (2 \pi)^4} 
\, {{\tilde p}^\mu {\tilde p}^\nu \over
{\tilde p}^4 } \, \sum_{k=0}^{N-1} \epsilon^{(k)} \, B_\mu^{(N\!-k)}(p) 
\, B_\nu^{(k)}(-p) \, ,
\label{action}
\end{equation}
with 
\begin{equation}
B_\mu^{(k)}={1 \over \sqrt{N}} 
\sum_{j=1}^N e^{2 \pi i \, {jk \over N}} \, {\rm Tr}\, A_\mu^{(j)} \, . 
\label{Bmu}
\end{equation}
Notice that ${B^{(k)}_\mu}^\dag = B^{(N-k)}_\mu$ and, consequently,
$\epsilon^{(k)}=\epsilon^{(N-k)}$. The tree level kinetic term for the 
center gauge bosons remains diagonal in terms of the $B_\mu^{(k)}$'s. 
Thus we may write a  
one-loop corrected dispersion relation for the modes of $B^{(k)}_\mu$ 
polarized along the non-commutative directions, which read
\beq
E^2={\vec p}^2-g^2 n  {\epsilon^{(k)} \over \pi^2}\, 
{1 \over {\tilde p}^2} \, .
\eeq

Adjoint scalars, in case they are present, also get pole-like infrared  
contributions to their self-energy. The couplings of the adjoint scalars 
to the other fields can be obtained by dimensional reduction from a theory 
in $D>4$ dimensions and where the only adjoint bosons are the gauge 
fields. Using this, we obtain the contribution to the effective action
(see \cite{Armoni:2001uw} for a detailed derivation for the case
of type 0 D3-branes)
\begin{equation}
\Delta S' ={g^2 \over 4 \pi^2} \int {d^4 p \over (2 \pi)^4} \, {1 \over
{\tilde p}^2 } \sum_{i,j=1}^N M_{ij} \, \sum_{l=1}^{D-4}
{\rm Tr}\, \phi_l^{(i)}(p) {\rm Tr}\, \phi_l^{(j)}(-p) \, ,
\label{polescalar}
\end{equation}  
where $\phi_l$ denote the adjoint (real) scalars. 
The same diagonalization procedure 
applied to $\Delta S$ can by used for $\Delta S'$. Only adjoint bosons 
(vectors or scalars) get corrected at one loop. Bi-fundamental bosons are 
not corrected. The reason, clearly seen in 't Hooft double line notation, 
is that one cannot draw a non-planar graph for the vacuum polarization if 
the external legs are in the bi-fundamental representation. Note also that 
fermions (in any representation) do not get corrections.

We will now show that there is a remarkable relation between the sign
of $\epsilon^{(k)}$ and the sign of the (mass)$^2$ of the lowest state in 
the $k^{th}$ twisted sector. Namely, we will see that when $\epsilon^{(k)}$ 
is positive, indicating an instability in the non-commutative gauge theory, 
the associated twisted sector contains tachyons, while when the sign of 
$\epsilon^{(k)}$ is  negative the corresponding twisted sector is 
non-supersymmetric but does not contain tachyons. Finally, vanishing of
$\epsilon^{(k)}$ (absence of pole-like IR divergences for the related 
gauge theory sector) is correlated with the twist corresponding to the 
$k^{th}$ closed string twisted sector being supersymmetry-preserving.
Indeed, this is the case iff $k a_\alpha=0\, 
{\rm mod}\, N$ for some $\alpha$, so $\epsilon^{(k)}=0$ from 
(\ref{eigenvalue}).

In order to prove the previous statement, we turn now to analyze the closed 
string spectrum. In the $k^{th}$ twisted sector, worldsheet bosonic 
complex coordinates $Z^i$, $Z^{\ov i}$ have moddings $n+\theta_i$, 
$n-\theta_i$, with $n\in {\bf Z}$ and $\theta_i=kb_i/N$; worldsheet 
fermionic complex fields $\psi^i$, $\psi^{\ov i}$ have moddings 
$n+1/2+\theta_i$, $n+1/2-\theta_i$ in the NS sector and $n+\theta_i$, 
$n-\theta_i$ in the R sector. The spectrum of the string, as a whole, is 
unchanged under the redefinition
\begin{equation}
\theta_i\to \theta_i+k_i
\label{redef}
\end{equation}
with $k_i\in Z$ to preserve the modding of the oscillators, and $\sum_i 
k_i=$ even, to guarantee that the recover the same GSO projection. One 
may use this freedom to put the redefined $\theta_i$ in the range 
$(-1,1]$, as we assume henceforth. 

In R sectors, the zero point energy vanishes and the (mass)$^2$ of 
the corresponding spacetime states is non-negative. Therefore, and due 
to level-matching, all states in the NS-R, R-NS and R-R sector are 
necessarily non-tachyonic. 

In the NS-NS sector, the lightest spacetime states in the string tower are 
potentially tachyonic. When $\theta_i$ are in the range $(-1,1]$, the 
states which can become tachyonic may be described as follows. Consider 
say the left moving sector and define a `groundstate' $|0\rangle$ to be 
annihilated by all bosonic oscillator with positive modding, and by the 
fermionic oscillators $\psi^i_{n+1/2+\theta_i}$ and $\psi^{\ov 
i}_{n+1/2-\theta_i}$ for $n\geq0$
\begin{equation}
\psi^i_{n+1/2+\theta_i}|0\rangle = 0 \quad ; \quad
\psi^{\ov i}_{n+1/2-\theta_i}|0\rangle = 0 \quad ; \quad \forall n\geq 0
\end{equation}
This coincides with the usual vacuum of the NS sector for small 
$\theta_i$. When some $|\theta_i|>1/2$ it is not the usual vacuum, since 
it is annihilated by some fermion oscillator with negative modding. 
However, it will be useful to use this state throughout the range 
$\theta_i\in (-1,1]$. The energy of this groundstate is given by 
the zero point energy contribution $E_0=\frac 12 (|\theta_1|+|\theta_2|+
|\theta_3|-1)$, 
and this state is  odd under $(-)^F$. In order to describe 
the lightest states for positive and negative $\theta_i$'s in the range 
$(-1,1]$ in a unified way, let us define the operators 
\begin{eqnarray}
& \Psi^i_{-1/2+|\theta_i|} = \psi^i_{-1/2+\theta^i} \quad \quad & {\rm 
for} \; \theta_i>0 \nonumber \\ 
& \Psi^i_{-1/2+|\theta_i|} = \psi^{\ov i}_{-1/2-\theta^i} \quad \quad & 
{\rm for} \; \theta_i<0 
\end{eqnarray}
The modding of the oscillator $\Psi^i_{-1/2+|\theta_i|}$ is therefore 
$-1/2+|\theta_i|$ for any sign of $\theta_i$. The lightest states 
surviving the GSO projection in the left sector are
\begin{eqnarray}
& \Psi^1_{-1/2+|\theta_1|}|0\rangle & 
\alpha' m_1^2= |\theta_2|+|\theta_3|-|\theta_1| \nonumber \\
& \Psi^2_{-1/2+|\theta_2|}|0\rangle & 
\alpha' m_2^2= |\theta_3|+|\theta_1|-|\theta_2|  \nonumber\\
& \Psi^3_{-1/2+|\theta_3|}|0\rangle & 
\alpha' m_3^2= |\theta_1|+|\theta_2|-|\theta_3|  \nonumber \\
&\Psi^1_{-1/2+|\theta_1|} \Psi^2_{-1/2+|\theta_2|}
\Psi^3_{-1/2+|\theta_3|}|0\rangle & 
\alpha'm^2_4= 2-\sum_{i=1}^3 |\theta_i|
\label{masses}
\end{eqnarray}
In the range $\theta_i\in(-1,1]$ no other state can become tachyonic.
Due to level-matching, the masses of the lowest states in the NS-NS sector 
will be twice these ones.

The freedom (\ref{redef}) allows to set the parameters $\theta_i$
not only in the range $(-1,1]$, but also all $\geq 0$ or all
$<0$. Having into account that $\theta_i=kb_i/N$ and the expression
of $b_i$ in terms of $a_\alpha$ (\ref{ba}), 
we see that (\ref{redef}) is equivalent to 
the freedom in defining the parameters $k a_\alpha$ modulo $N$. 
Using these two facts, it is straightforward to show 
\begin{eqnarray}
\alpha'm^2_1= 2 \epsilon {k a_1 \over N} +r_1& , \,\,\,\,\, &
\alpha'm^2_2= 2 \epsilon {k a_2 \over N} +r_2 \, ,\nonumber \\
\alpha'm^2_3= 2 \epsilon {k a_3 \over N} +r_3& , \,\,\,\,\, &
\alpha'm^2_4= 2 + 2 \epsilon k{a_4 \over N}-(r_1+r_2+r_3) \, ,
\end{eqnarray}
with $r_i$ certain even integers, and $\epsilon=1$ when all 
$\theta_i\geq 0$ and $\epsilon=-1$ when all $\theta_i< 0$. 
Using (\ref{eigenvalue}) we can rewrite the eigenvalues of $M$ as
\begin{equation}
\epsilon^{(k)}=-16 \prod_\alpha\sin {\pi \alpha' m^2_\alpha \over 2}  \, .
\label{eigenM}
\end{equation}

We can now compare the sign of the IR pole coefficient $\epsilon^{(k)}$
and the sign of the (mass)$^2$ of the lightest state in the k$^{th}$
closed string twisted sector. From (\ref{masses}) we observe that 
$-1 \leq \alpha' m^2_\alpha \leq 2$, and hence the sign of
$\sin {\pi \alpha'm^2_\alpha \over 2}$ will coincide 
with that of $m^2_\alpha$. If at most one of the masses 
$m^2_\alpha$ could be negative, we would have obtained 
the conjectured statement: the sign of $\epsilon^{(k)}$ is positive if the 
associated closed string twisted sector contains a tachyon, and negative 
if the closed string sector is non-supersymmetric but does not contain 
tachyons. In order to show that at most one of the $m^2_\alpha$
can be negative, it is convenient to label (without loss of generality)
the parameters $\theta_i$ such that 
\beq
|\theta_1| \leq |\theta_2| \leq |\theta_3| \, .
\eeq
It is clear then that only $m^2_3$ and $m^2_4$ could be negative.
$m^2_3<0$ implies that $|\theta_1|+|\theta_2|<|\theta_3|$, while
$m^2_4<0$ implies that $|\theta_1|+|\theta_2|>2-|\theta_3|$. Both
conditions can not be met at the same time, since they would
mean that $1<|\theta_3|$, which contradicts the fact that the
parameters $\theta_i$ has been chosen in the interval $(-1,1]$.
Thus we obtain the remarkable result that the presence or not of 
instabilities of the non-commutative quantum field theory is correlated 
with the presence or not of tachyons in the closed string spectrum of the 
string theory realization of the configuration.

\medskip

The fact that the non-commutative field theory instabilities contain some 
information about the closed string sector of the string theory 
configuration can be put to a slightly stronger test. In fact we can 
consider different quantum field theories associated to the same orbifold 
singularity, by considering D3-branes in representations of $Z_N$ other 
than the regular one. The resulting field theories have a field content 
similar to the above with the only difference that the ranks of the gauge 
factors are arbitrary $U(n_i)$ \footnote{Cancellation of RR twisted 
tadpoles implies some constraints on this numbers which may be understood 
in terms of cancellation of cubic non-abelian anomalies in the usual 
commutative situation \cite{Leigh:1998hj}. We consider such 
constraints to be satisfied, since they do not modify our argument.}, and 
can even vanish (so the corresponding gauge factor is absent).
If the instabilities of the non-commutative field theories are 
indeed correlated with the existence of tachyons in closed string 
spectrum, all field theories obtained from the same orbifold singularity 
(i.e. described by the same quiver diagram) should lead to the same signs 
for the corresponding eigenvalues.

It is easy to see that the structure of the IR poles in this more general 
situation is still given by (\ref{pole}), (\ref{polematrix}), with all the 
dependence on the arbitrary ranks $n_i$ included in the normalization of 
${\rm Tr}\, A_\mu^{(i)}$. For this it is important to realize that the 
gauge couplings for different gauge factors are still equal in this more 
general situation. Therefore, the diagonalization of (\ref{polematrix}) 
proceeds exactly as above, via linear combinations of the form \eqref{Bmu}, 
with the proviso that if some $n_i=0$ the corresponding gauge boson 
disappears from the linear combination. The diagonalization hence leads 
to the same eigenvalues $\epsilon^{(k)}$. Hence the sign of the one-loop 
pole contribution is again correlated with the presence or not of tachyons 
in the closed string twisted sector. Note that in the basis diagonalizing 
the one-loop contribution, the tree level kinetic term is not diagonal.
However, in the generic case of non-zero $n_i$'s, the sign of 
the $\epsilon_k$'s correlated with field theory instabilities, since the 
latter occur at low momenta where the pole dominates the effective action.
When some $n_i=0$, the fields $B_k$ are not all independent, and it is 
linear combinations of the $\epsilon_k$ that determine the sign of the
pole for the independent fields, as in the example below.

To verify that our general recipe is valid even when some $n_i$ vanish,
it is interesting to further explore the extreme situation where all 
$n_i$'s vanish except for one. In the non-supersymmetric case (no 
$a_\alpha=0$), the quantum field theory reduces to pure $U(n_j)$ 
Yang-Mills, coupled to adjoint complex scalars if some $b_\beta=0$. 
In this situation, 
\begin{equation}
B_\mu^{(k)}= \frac{1}{\sqrt N} e^{2 \pi i \, {jk \over N}} 
{\rm Tr}\, A_\mu^{(j)}
\end{equation}
and our general formula (\ref{action}) would give
\begin{equation}
\Delta S = {g^2 \over 2 \pi^2} \int {d^4 p \over (2 \pi)^4}
\; {{\tilde p}^\mu {\tilde p}^\nu \over
{\tilde p}^4 } \, \frac 1N \, \sum_{k=0}^{N-1} \epsilon^{(k)} 
{\rm Tr}\, A_\mu^{(j)}(p)\, {\rm Tr}\, A_\nu^{(j)}(-p)
\label{action2}
\end{equation}
The coefficient should equal the contribution directly computed from the 
spectrum of gauge bosons and adjoint scalars. This requires a sum rule for 
the eigenvalues, which is happily satisfied
\begin{eqnarray}
& \frac 1N \, \sum_{k=1}^N \epsilon^{(k)} \, =\, \frac 1N \, \sum_{k=1}^N 
\, 2\left(1-\sum_\alpha \cos {2 \pi k a_\alpha \over N}
+\sum_\beta \cos {2 \pi k b_\beta \over N}\right) \, \nonumber \\
& = 2+2\sum_{\beta=1}^3 \delta_{b_\beta,0}
\end{eqnarray}
since the later is precisely the contribution to the pole from gauge 
bosons and adjoint scalars (if present).

\section{String Interpretation}
\noindent

In the previous section we analyzed both the one-loop field theory
spectrum and the tree-level spectrum of the closed string theory. We
found that they are related. 
In this section we will further explore the string
origin of (\ref{action}).

Let us consider again D3-branes in the regular representation of a 
type IIB $C^3/Z_N$ orbifold.
The contribution to the effective action coming from the 
pole-like infra-red divergences (\ref{action}) is not gauge 
invariant. However it can be completed to produce a gauge
invariant expression \cite{VanRaamsdonk:2001jd,Armoni:2001uw,Kiem:2001dm}
\beq
\Delta S = {1 \over 2 \pi^2}  \sum_{k=0}^{N-1} \int {d^4 p \over (2 \pi)^4}
\; {\epsilon^{(k)} \over
{\tilde p}^4}\; W^{(N\!-k)}(p)\, W^{(k)}(-p) \, ,
\label{invaction}
\eeq
with 
\beq
W^{(k)} = {1 \over {\sqrt N}} \sum_{j=1}^N e^{2 \pi i \, {jk \over N}} \, 
{\widetilde W}^{(j)} \, ,
\label{wtwisted}
\eeq
and ${\widetilde W}^{(j)}$ denoting the simplest gauge invariant open 
Wilson line operators \cite{Ishibashi:2000hs,Gross:2000ba}
\beq
{\widetilde W}^{(j)}(p)\, = \, {\rm Tr} \int d^4 x \; P_\ast 
\left(e^{i\, g \int_0^1 d \sigma 
\, {\tilde p}^\mu A^{(j)}_{\mu}(x+{\tilde p}\, \sigma)} \right) 
\ast e^{i p x} \, .
\label{w}
\eeq
Notice that the terms ${\rm Tr}\,1$ in the expansion of the exponential
cancel in $W^{(k)}$ for $k \neq 0$, while for $k=0$ we have 
$\epsilon^{(0)}=0$. Thus, the leading term in the expansion of 
(\ref{invaction}) reproduces (\ref{action}).

In the context of D-branes in B-field backgrounds, it has been shown 
that closed string modes couple naturally to straight open Wilson line 
operators \cite{Das:2001ur,Okawa:2001sh,Liu:2001ps}.
The closed strings in the $k^{th}$ twisted sector couple to gauge invariant
operators formed out of fields in the adjoint of the $j^{th}$
gauge group as \cite{Douglas:1996sw}
\beq
{\rm Tr} (\gamma_k \lambda_j) \, \phi_k\, {\cal O}_j= 
e^{2 \pi i \, {jk \over N}} \phi_k \, {\cal O}_j \, ,
\label{twisted}
\eeq
where $\gamma_k$ is the Chan-Paton matrix associated to the $k^{th}$ twist 
and $\lambda_j$ projects on the Chan-Paton indices of the $j^{th}$
gauge group factor. Using this 
we observe that the operators $W^{(k)}$ defined in (\ref{wtwisted}) 
will couple to closed strings in the $k^{th}$ twisted sector.
Therefore the term in the gauge theory effective action 
responsible for the instability (\ref{invaction}) looks formally like 
a closed string exchange between D-branes.
The closed string modes that can contribute to (\ref{invaction})
must have zero spin since $W^{(k)}$ carry no spin \footnote{More in 
general, the closed string modes that can couple to $W^{(k)}$ must be 
singlets under the unbroken subgroup of the Lorentz group.}. 
We have seen in the previous section that tachyonic modes have 
always zero spin. It is interesting to notice that they are the 
first candidates to couple to $W^{(k)}$. 
Motivated by this analogy, it is natural to wonder whether
$1/{\tilde p}^4$ in (\ref{invaction}) can be related in some way  
to a closed string propagator. We will now analyze this question, 
generalizing the approach developed in \cite{Armoni:2001uw} 
for type 0 D3-branes.

Spin zero closed string modes
will couple to a series of field theory operators weighted by appropriate
$\alpha'$ powers. At leading order in $\alpha'$ they generically couple
to the brane tension. In the absence of an expectation value of the B-field, 
the field theory operator associated to the tension of a set of D3-branes 
is just ${\rm Tr}\ 1$. In the presence of a B-field background, 
${\rm Tr}\ 1$ is promoted to the simplest straight open Wilson line operator.
Let $\phi$ be an spin zero closed string mode belonging to the $k^{th}$ 
twisted sector of the orbifold $C^3/Z_N$. At leading $\alpha'$ order, 
$\phi$ will couple to the field theory operator
\beq
{\cal O}(p)={c_\phi \over (2 \pi \alpha')^2} \, W^{(k)}(p) \, ,
\label{op}
\eeq
where $W^{(k)}$ is given by (\ref{wtwisted}) and $c_\phi$ is a numerical
factor which can be extracted from the disk amplitude with an insertion
of the closed string mode $\phi$ and no open string insertions. 
At sub-leading order in $\alpha'$,
$\phi$ will couple to open Wilson line operators with the insertion of
non-trivial spin zero field theory operators, for example $F^2$.
We are interested in giving a string theory interpretation to 
(\ref{invaction}), therefore from now on we will concentrate on the coupling 
of $\phi$ to ${\cal O}$ as in (\ref{op}). 

In the presence of an expectation value of the B-field, open and closed
string modes neither perceive the same space-time metric, nor string 
coupling constant \cite{Seiberg:1999vs}. The following relation holds
\beq
G^{-1}+{\theta \over 2 \pi \alpha'}= {1 \over g + 2 \pi \alpha' B} 
\;\;\;\;\;\; , \;\;
G_s=g_s \left({{\rm det} G \over {\rm det}g}\right)^{1 \over 4} \, ,
\label{ocl}
\eeq
where $g$ and $g_s$ ($G$ and $G_s$) denote the closed (open) string 
metric and coupling constant respectively. The coupling of $\phi$ to 
the field theory operator ${\cal O}$ is described by the action
\beq
S={k_D \over 2 \pi G_s} \int {d^4 p \over (2 \pi)^4} \, 
\sqrt{{\rm det} G} 
\, \phi(-p) \, {\cal O}(p) \, ,
\eeq
where $D$ is the number of dimensions where the twisted field 
$\phi$ can propagate. In the generic $C^3/Z_N$ orbifold, we have $D=4$.
However, depending on the particular orbifold model and twisted sector
we can also have $D=6,8,10$. $k_D$ is the gravitational coupling 
constant in $D$-dimensions
\beq
k_D^2 \sim g_s^2 {\alpha'}^{D-2 \over 2} \, .
\eeq

The contribution to the field theory effective action from the
exchange of $\phi$ between two D3-branes separated by a distance
$r$ will be
\beq
\Delta S= {k_D^2 \over (2 \pi G_s)^2} \int 
{d^4p \over (2 \pi)^4} \, {\rm det}G \, {\cal O}^{\dag}(p) \, {\cal O}(-p) 
\left( {1 \over \sqrt{{\rm det} g}} \int {d^d p_\perp \over
(2 \pi)^d} {e^{i p_{\perp} r} \over p_{\perp}^2+M^2} \right) \, ,
\label{exchange}
\eeq
where $p$ and $p_\perp$ denote the components of the momentum along and
transversal to the D3-brane respectively. The expression in brackets 
is the closed string propagator in $d=D-4$ dimensions. The above
expression \eqref{exchange} is valid for NS-NS fields. R-R exchange
results with an effective action with an opposite sign. $M^2$ acts 
as an effective mass term for the propagation in the transverse 
dimensions
\beq
M^2=m^2-g^{\mu \nu} p_\mu p_\nu=m^2 -p^2 + 
{{\tilde p}^2 \over (2 \pi \alpha')^2}  \, ,
\eeq
with $m^2$ being the mass of $\phi$ in D-dimensions. We have used 
$g^{-1}=G^{-1}\!-\! \theta G \theta/(2 \pi \alpha')^2$, and denoted 
$p^2=p\!\cdot\! G^{-1}\!\cdot\! p$ and ${\tilde p}^2=-{\tilde p}
\! \cdot \!G \! \cdot \!{\tilde p}$.
The non-commutative field theory 
limit consists in sending $\alpha' \rightarrow 0$ while keeping
$G$, $\theta$ and $g_{YM}^2=2 \pi G_s$ fixed. It is in this limit 
where we want to analyze (\ref{exchange}). Notice that when $B=0$,
the leading $\alpha'$ effect is related to $m^2\sim 1/\alpha'$.
However when the field theory limit is taken at fixed non-zero $\theta$, 
$m^2$ becomes a sub-leading effect if the closed string mode carry 
momentum along the non-commutative directions 
\cite{Arcioni:2000bz,Okawa:2001sh}. In the following we will
neglect the dependence of $M^2$ on $p^2$, suppressed by two powers of 
$\alpha'$ with respect to ${\tilde p}^2$, but will retain its dependence 
on $m^2$.

It is convenient to introduce $y=2 \pi \alpha' p_\perp$, 
$u=r/(2 \pi \alpha')$ and set $G=\eta$, the Minkowski metric.
Using (\ref{ocl}) and (\ref{op}) we obtain\footnote{See \cite{Murakami:2002yd}
for an alternative derivation of an analogous expression
in the context of bosonic string theory.}
\beq
\Delta S=  \int {d^4 p \over (2 \pi)^4} \, 
W^{(N-k)}(p) W^{(k)}(-p) \, f({\tilde p},u) \, 
\label{staction}
\eeq
with
\beq
f({\tilde p},u)={|c_\phi|^2 \, k_D^2 \over (2 \pi g_s)^2 
(2 \pi \alpha')^{D-2}} \int {d^d y \over
(2 \pi)^d} {e^{i y u} \over y^2+{\tilde p}^2+(2 \pi \alpha' m)^2} \, .
\eeq
The fraction in front of the integral is independent of $g_s$
and proportional to $\alpha'^{1-D/2}$. The integral is finite when
$\alpha' \rightarrow 0$. Thus $f({\tilde p},u)$ diverges when we send 
$\alpha' \rightarrow 0$. We could however define a finite contribution
by expanding the integral in powers of $2 \pi \alpha' m$ to order 
$\alpha'^{D/2-1}$
\beq
f({\tilde p},u)|_{{\cal O}(\alpha'^0)}\, \sim \, |c_\phi|^2 \, 
(\alpha' m^2)^{{D \over 2}-1} \int {d^d y \over
(2 \pi)^d} {e^{i y u} \over (y^2+{\tilde p}^2)^{D \over 2}} \, .
\label{ffin}
\eeq
With this prescription (\ref{staction}) defines a finite contribution
to the field theory 1-loop effective action.
The parameter $u$, measuring the separation between the two D3-branes,
has the field theory interpretation of a mass for the degrees of freedom 
circulating in the loop. In the non-planar field theory loops the scale  
$1/{\tilde p}$ acts as ultraviolet cutoff. Thus when ${\tilde p} u >1$
the contribution of non-planar graphs is strongly suppressed. We observe
that this pattern is reproduced by (\ref{ffin}). The parameter $u$ was 
introduced in order to have a well-defined closed string 
propagator. Now that we have derived from $f({\tilde p},u)$ an expression 
candidate to have a field theory interpretation, we can let 
$u \rightarrow 0$. The result is 
\beq
f({\tilde p},0)|_{{\cal O}(\alpha'^0)}\, \sim \, { |c_\phi|^2 \, 
(\alpha' m^2)^{{D \over 2}-1} \over {\tilde p}^4} \, .
\label{stprop}
\eeq
Hence the term $1/{\tilde p}^4$, source of the pole 
like IR divergent behavior, can be traced back to a closed string
propagator independently of the dimension D in which the closed
string modes live. This does not mean that the decoupling 
limit fails in the non-commutative case, since the IR singularities 
do not have kinetic part. Therefore they do not force the introduction 
of new degrees of freedom.

We have seen that considering the exchange of a 
single closed string mode between D-branes, we are able to reproduce 
the field theory effective action (\ref{invaction}). However this is
achieved {\it up to an overall factor}. It is crucial to 
notice that all closed string modes (of spin zero or spin zero 
combinations of them) contribute as in (\ref{staction}) and (\ref{stprop}).
Indeed by open/closed channel duality the contribution from the lowest 
modes in the open string tower are mapped to the contribution from
an infinite number of closed string oscillators. Thus while the
functional dependence on $\tilde p$ can be related to a single closed 
string propagator, the numerical factor $\epsilon^{(k)}$
is a collective effect of the whole closed string tower. 

The surprising fact is that, although it includes the contribution from all
closed string modes, $\epsilon^{(k)}$ is at the same time determined
by the lowest states in the tower. These two facts can be reconciled in 
the following heuristic way. Modes in the NS-NS sector contribute with 
different sign to those in the R-R sector to the exchange between 
D-branes.
In twisted sectors corresponding to a supersymmetry-preserving twist, 
both towers must cancel since we know that in this case 
$\epsilon^{(k)}=0$. This is actually easy to see 
also from the string approach: The Wilson line operators $W^{(k)}$ 
generalize the coupling of the closed string modes to the brane tension in 
the case of non-zero $B$-field; Thus, 
for the exchange of states in twisted sectors corresponding to 
supersymmetry-preserving twists
the condition of no force between the branes 
\footnote{In the evaluation of the annulus 
diagram, sectors with a supersymmetry-preserving twist
behave effectively as supersymmetric.} implies that 
$\epsilon^{(k)}=0$ (see below). In particular, the untwisted sector 
of the orbifold models 
corresponds to the trivial twist, which is supersymmetry-preserving,
and hence $\epsilon^{(0)}=0$. This 
suggests that, 
for twisted sectors associated to non-supersymmetric twists,
$\epsilon^{(k)}$ can be 
considered a measurement of the misalignment between the NS-NS and R-R 
towers. On the other hand, the expression of $\epsilon^{(k)}$ in terms of 
the masses of the lowest states in the NS-NS tower (\ref{eigenM}) can be 
also seen as a measure of the misalignment between both towers, since the 
lightest states in the R-R sector are massless. At any rate, we are 
uncovering a new relation fulfilled by the closed string spectrum, 
reminiscent of some results based in modular invariance (in that they 
related properties of the whole string tower with properties of the 
lightest modes), which would be very interesting to re-derive from first 
principles.

The gauge invariant effective action (\ref{invaction}) is also valid
in the case we are considering D3-branes in a generic the representation
of the orbifold group. The only difference then is that
the Wilson line operators ${\widetilde W}^{(j)}$ should be redefined
as
\beq
{\widetilde W}^{(j)}(p)\, = \, {\rm Tr} \int d^4 x \; P_\ast 
\left(e^{i\, g \int_0^1 d \sigma 
\, {\tilde p}^\mu A^{(j)}_{\mu}(x+{\tilde p}\, \sigma)} -1 \right) 
\ast e^{i p x} \, .
\label{wd}
\eeq
Contrary to the case of regular D3-branes, the components ${\rm Tr}_j 1=n_j$ 
of ${\widetilde W}^{(j)}$ do not cancel in the definition
(\ref{wtwisted}). Therefore, in order that (\ref{invaction}) does not
contain spurious terms, they have to be explicitly subtracted.

Notice that the closed string modes couple to the complete Wilson 
line operators. This does not affect our string derivation of the field 
theory effective action because it was only valid at non-zero momentum
in the non-commutative directions. Subtracting or not the constant 
term $n_j$ from ${\widetilde W}^{(j)}$
affects only its zero momentum component. 

However, the string derivation provides an interpretation for the 
terms proportional to $n_j$. As we have seen these terms 
correspond to the coupling of the closed string modes to the brane 
tension. Therefore channel duality implies that by considering the 
exchange of all closed string modes, in the $\alpha' \rightarrow 0$ limit, 
we obtain the contribution of one-loop vacuum diagrams to the field theory 
effective action
\beq
\Gamma= -{1 \over 2} \,V\,{\rm Tr} (-1)^F \int {d^4 p \over (2 \pi)^4} \,
{\rm log} \, p^2 \, .
\label{vac}
\eeq
where $V$ represents the infinite 4-dimensional volume. We will now
show explicitly that (\ref{vac}) coincides with (\ref{invaction}) when the
Wilson line operators are replaced by their constant component.
Substituting ${\widetilde W}^{(j)} \rightarrow (2 \pi)^4 n_j
\delta^{(4)}(p)$ in (\ref{invaction}) we obtain
\beq
{1 \over 2 \pi^2} \Lambda^4 \left( \sum_{k=0}^{N-1} 
\epsilon^{(k)} v^{(N-k)} v^{(k)} \right) \int {d^4p \over (2 \pi)^4}
\left((2 \pi)^4 \delta^{(4)}(p)\right)^2 \, ,
\label{actionvac}
\eeq
where we have defined $v^{(k)}={1 \over \sqrt{N}} \sum_{j=1}^N 
e^{2 \pi i kj \over N} n_j$. 
The delta function that substitute the Wilson line operators set $p$ 
and thus ${\tilde p}$ to zero. Therefore the factor $1/{\tilde p}^4$ 
in (\ref{invaction}) will give rise to a divergence. Since we want to
compare (\ref{actionvac}) with the planar contribution coming from 
vacuum loops, we have replaced in it $1/{\tilde p}^4|_{{\tilde p}=0}$ 
by $\Lambda^4$. We interpret $\Lambda$ as an UV regulator.

The integral in (\ref{vac}) is quartically divergent.
It can be evaluated by introducing an UV cutoff
\beq
-{1 \over 2} \int {d^4 p \over (2 \pi)^4} \,{\rm log} \, p^2
\rightarrow {1 \over 8 \pi^2} \int_0^\infty {dt \over t^3} 
e^{-{1 \over 4 \Lambda^2 t}} = {1 \over 2 \pi^2} \Lambda^4 \, .
\eeq
The cutoff has been chosen by analogy to how $1/{\tilde p}$ regulates the
UV region of non-planar integrals. This is necessary in order to make the
substitution $1/{\tilde p}\rightarrow \Lambda $ in (\ref{actionvac})
meaningful. We also have
\beq
\sum_{k=0}^{N-1} \epsilon^{(k)} v^{(N-k)} v^{(k)} = 
\sum_{i,j=1}^N M_{ij} n_i  n_j  = {\rm Tr} (-1)^F \, ,
\eeq
where the matrix $M_{ij}$ is given in (\ref{polematrix}).
Finally the integral in (\ref{actionvac}) is just $\int d^4 x
=V$ written in momentum space. This shows that (\ref{actionvac})
exactly reproduces the contribution from the field theory
1-loop vacuum diagrams, and how this depend on the
coefficients $\epsilon^{(k)}$. In particular $\epsilon^{(k)}$
govern in a similar way the term in the effective
action responsible for the instabilities (\ref{invaction}) and 
the contribution from the vacuum loops.

We would like to end this section with a comment on the cases  
with scalars in the adjoint representation. Adjoint scalars, as
gauge bosons, get pole-like IR corrections to their self-energy. 
This can be incorporated in the string derivation by just 
generalizing the definition of the Wilson line operators to 
\beq
{\widetilde W}^{(j)}(p)\, = \, {\rm Tr} \int d^4 x \; P_\ast 
\left(e^{i\, g \int_0^1 d \sigma \big(
\, {\tilde p}^\mu A^{(j)}_{\mu}(x+{\tilde p}\, \sigma) + 
y_i \phi_i^{(j)}(x+ {\tilde p} \sigma) \big)}\right) 
\ast e^{i p x} \, ,
\label{gw}
\eeq
where $y_i$ label as before the momentum in the directions
transverse to the D3-branes and $\phi_i$ denote the adjoint
(real) scalars. Indeed, it has been shown 
\cite{Okawa:2001sh,Liu:2001ps} that (\ref{gw})
are the field theory operators to which closed string modes 
naturally couple when adjoint scalars are present.
For a detailed derivation of how (\ref{gw}) gives rise to the
pole-like IR corrections to the self-energy of adjoint scalars
on type 0 D3-branes see \cite{Armoni:2001uw}. 
The discussion there generalizes straightforwardly to the more
generic orbifold models considered in this paper.

\section{Discussion}
\noindent

String theories without space-time supersymmetry tend to have tachyons in 
the spectrum. This statement can be done more precise by analyzing the 
torus partition function
\beq
Z=\int {d^2 \tau \over \tau_2} 
{\rm Tr}  (-)^F q^{L_0}{\bar q}^{{\bar L}_0} \, ,
\label{torus}
\eeq
where $q=e^{2 \pi i \tau}$ and $\tau=\tau_1 + i \tau_2$ is the complex 
modulus of the torus. 
This integral can have both UV ($\tau_2\rightarrow 0$) and IR 
($\tau_2\rightarrow \infty$) divergences. It is easy to see that an IR 
divergence can only appear if there are tachyons in the spectrum. The 
trace in (\ref{torus}) reduces in the UV to the supertrace over the number 
of degrees of freedom. Modular invariance of (\ref{torus}) relates the UV 
and IR regimes. As a result, the absence of closed string tachyons 
requires that the density of space-time bosons and fermions 
cancel to a great accuracy, although both quantities will generically 
grow exponentially with the energy \cite{Kutasov:1990sv}.
String theories without this sort of ``asymptotic supersymmetry'' will 
have tachyons in the spectrum. This provides a very precise relation
between tachyons and absence of supersymmetry in string theory.

String theory regularizes the UV region of the torus partition function 
by restricting the integration to the fundamental domain of the modular 
group. But possible IR divergences remain. String theory translates the 
possible divergence on vacuum diagrams because of the growing density of 
states into an IR divergence. This is a particular case of the general 
situation: all divergences in string theory have an IR interpretation. 
Therefore, contrary to UV divergences which are susceptible of being
regularized, they have a physical meaning and can not be discarded.
 
The extremely remarkable thing is that a version of all these features 
seems to survive in non-commutative field theories, without the inclusion 
of gravity. First, non-planar graphs of non-commutative field theories 
translate UV into IR behavior. In particular, they transform potential
UV divergences into IR divergences. This phenomenon however does not
take place in the planar sector of the theory. As an interesting example, 
we have seen that the quartic divergences of the field theory vacuum 
energy do have an infrared translation in (\ref{invaction}).
Second, absence of supersymmetry implies strong modifications of the spectrum 
of the theory at low momentum. This is specially neat for non-commutative 
gauge theories. Indeed, the effective action (\ref{invaction}) implies the 
presence of unstable modes in one to one correspondence with closed string 
tachyons for those gauge theories which can be derived as limits of 
string theory. Although a deeper understanding is lacking 
of why these phenomena are present in non-commutative field 
theories, it agrees with the proposal that non-commutative 
geometry is able to encode in an effective way crucial aspects of quantum 
gravity \cite{Doplicher:tu}.

\medskip

We would like to point out that the correlation between the existence (or
not) of instabilities in the non-commutative field theory and the
existence (or not) of tachyons in the closed sector of the underlying
string theory, seems to be quite general, and not restricted to the family
of non-supersymmetric orbifold theories we have been considering. In fact,
it is possible to find such a relation in particular examples of
non-commutative field theories arising on the volume of stable non-BPS
branes in supersymmetric theories. In particular, one may test the issue
by considering the configurations studied in \cite{Sen:1998ex}, namely
compactification of type IIB theory on the orbifold limit $T^4/Z_2$ of K3,
with a stack of $n$ (stable at short radii) non-BPS D4-branes wrapped on a
1-cycle passing through fixed points. The field theory on the 4d
non-compact directions of the D-branes has gauge group $U(n)$, four
Majorana fermions in the adjoint and one adjoint complex scalar. Turning
on a NSNS 2-form field in the non-compact directions makes this theory
non-commutative; the field content reveals that there is a one-loop
correction to the mass of gauge bosons, but no instability. This agrees
with our claim, since the closed string sector does not contain tachyons.

We should stress that although the closed string sector is
supersymmetric, the one-loop correction does not vanish, as was the case
for D3-branes at orbifold singularities. We blame this on the non-BPS
character of the corresponding D-branes, while for D3-branes at orbifold
singularities, the D3-branes were in a sense, still BPS states of the
theory before orbifolding.

\medskip

${\cal N}=4$ non-commutative $U(n)$ theory at finite temperature was 
analyzed in \cite{Landsteiner:2001ky}. Temperature breaks supersymmetry but
at the same time acts as an UV regulator. As a consequence this theory 
presents a regularized version of UV/IR mixing. The strength of
UV/IR mixing effects grows with the temperature, such that for
$T$ above a critical temperature $T_c\sim 1/\sqrt{g \theta}$ 
collective excitations of tachyonic nature appear in the system.
An important characteristic of this tachyonic branch is that $E^2(p)
\rightarrow 0$ for low momenta. Namely, contrary to the behavior in
purely non-supersymmetric theories, UV/IR effects give rise to
instabilities without the appearance of IR divergences.
Let us consider the string embedding of this theory, namely,
a D3-brane in Type IIB at finite temperature. It is well known that
string theory becomes unstable at the Haggedorn temperature, 
$T_H \sim 1/\sqrt{\alpha'}$. Following the spirit of this paper it is 
tempting to try to relate both instabilities, and in particular both
critical temperatures. We find however a problem. In the 
$\alpha' \rightarrow 0$ limit we have $T_H \rightarrow \infty > T_c$. 
This suggests that the relation between non-commutative instabilities 
and closed string tachyons could make sense 
only for field theory instabilities where $E^2<0$ as $p\rightarrow 0$. 
Indeed, when the field theory temperature is sent to infinity,
an IR divergence is generated \cite{Landsteiner:2001ky}.

It is also interesting to ask to what extent the non-commutative 
nature of the field theory is relevant for the connection with closed 
string tachyons. Non-commutative gauge theories exists only for
unitary gauge groups and matter in the fundamental, bifundamental and
adjoint representations. We have seen in what remarkable way 
bifundamental and adjoint degrees of freedom can be related with
closed string tachyons for theories on D3-branes. Fundamental matter,
if present, does not contribute to this relation since it does not
give rise to non-planar diagrams. We could ask then whether this relation
can be extended to gauge groups and matter content which do not
admit a non-commutative version\footnote{Using a generalization of
the Seiberg-Witten map, it has been proposed a way of defining 
non-commutative theories with arbitrary gauge group and matter
content \cite{Jurco:2000dx}. However a string realization of
this setup does not exist.}. In fact, one may consider quantities like 
the number of bosonic minus fermionic degrees of freedom, weighted in a 
suitable way to account for gauge quantum numbers (say, matrices like 
(\ref{polematrix})) in any field theory on D3-branes, and try to correlate 
the corresponding signs with closed string tachyons. Interesting
examples to analyze would be cases including orientifold projections. 
We leave this as an open question for further study.
 
\medskip

Let us come back to the particular setup of orbifold singularities, and 
discuss the relation between our statements and other approaches in the 
literature.
The fate of twisted closed string tachyons of $C/Z_N$ and $C^2/Z_N$ 
(N odd) non-supersymmetric orbifolds has been studied
\cite{Adams:2001sv}. It was argued that 
the end point of the condensation will generically be flat space. 
Information about this process can be obtained by placing 
D-branes probes at the orbifold point. The Higgs branch of the probe gauge
theory, associated to expectation values of the bifundamental fields, 
reproduces the geometry of the orbifold. As the tachyon starts condensing
new tree level couplings appear in the gauge theory. Among them are
masses for the bifundamental fields which, due to coming from twisted 
sectors, must add up to zero (see (\ref{twisted})). Thus at least one 
bifundamental must become unstable. As a result the Higgs branch is 
modified and represents a geometry in which the tip of the cone has been 
smoothed out. All the analysis is done at the classical level, keeping
$\alpha'$ finite.

Contrary to that, we 
consider non-commutative D3-branes at the tip of an $C^3/Z_N$ orbifold in the
$\alpha' \rightarrow 0$ limit. In this situation the branes decouple from the 
gravity sector. The gauge theory instability is a 1-loop effect. In addition, 
the instability affects the adjoint fields and not the bi-fundamentals. 
Type 0 D3-branes \cite{Zarembo:1999hn,Tseytlin:1999ii} and D3-branes at the 
tip of non-supersymmetric orbifold singularities \cite{Adams:2001jb} 
has been also studied in the decoupling limit for the case of zero 
$B$-field. For models with a Coulomb branch, a Coleman-Weinberg effective 
potential for the diagonal components of the scalar
fields was calculated. This potential, associated with twisted operators 
under the orbifold action, showed an instability towards the separation 
of the branes. The instability could be considered as inherited from the 
original twisted string tachyons. However, the correspondence in this 
case was far from one to one. First it required
configurations with several branes. Then it was only manifest when there
was a Coulomb branch in the theory, representing dimensions along which
the branes could separate. Non-supersymmetric $C^3/Z_N$ models without adjoint
scalars do not exhibit these instabilities.

One could hope that the D3-brane models in the decoupling limit could
serve as toy models of the parent string theory instabilities.
In the case with zero $B$-field, as we have mentioned, the field theory 
dynamics is not rich enough to fulfill this expectation. The case in 
which the decoupling of gravity limit is taken at non-zero $B$ is
potentially different. The $B$-field allows the field theory to keep 
some stringy properties. It is natural then to wonder if 
the gauge theory instabilities will have a related fate to that of closed 
string tachyons. Notice that not only there is a one to one correspondence 
between both instabilities, but also both have a related origin in the 
appearance of divergences due to the absence of supersymmetry. We will not 
attempt to answer this question here.
Understanding of the dynamics of the non-commutative instability is 
beyond the scope of this paper. However we would like to add some 
comments. 

A non-commutative theory can be alternatively formulated in terms of
a matrix theory of infinite-dimensional matrices in an space-time 
with two dimensions less. In string language this means that 
a D$p$-brane in a B-field background possesses a non-zero density of
D$(p-2)$-branes dissolved on it. The non-commutative instability 
translates in this language into an instability of the smeared 
D$(p-2)$-brane distribution \cite{VanRaamsdonk:2001jd,Landsteiner:2001ky}.
The origin of the instability is that the forces between these smaller 
branes do not cancel in the non-supersymmetric case, which suggest that 
there is no stable vacuum \cite{VanRaamsdonk:2001jd}. 
This situation relates with the cases treated in this paper 
where there are fractional branes present. Namely branes fixed at the 
orbifold point, which do not have an analog in the flat supersymmetric 
theory. For D3-branes at orbifold singularities, the Chern-Simons coupling 
between the R-R two-forms $C^{(2)}_k$, for $k=0,..,N-1$, and the $B$-field
is proportional to \cite{Douglas:1996sw}
\beq
{\rm Tr}\, \gamma_k \, \int_4 C_k^{(2)} \wedge B=  
\left( \sum_j n_j e^{2 \pi i {jk \over N}} \right) B \, \int_2 C_k^{(2)} \, 
\label{cs}
\eeq
where we have used (\ref{twisted}) and the fact that $B$ belongs
to the untwisted sector. We observe that when there are
fractional branes present, i.e. the ranks of the $N$ gauge group 
factors do not all coincide, the $B$-field acts as a source for some
of the twisted R-R 2-forms. Hence the initial brane configurations contains 
a background charge which under the effective action (\ref{invaction})
tends to collapse, giving likely rise to no stable vacuum 
\cite{VanRaamsdonk:2001jd}.

A potentially different case is that of branes in the regular 
representation of the orbifold group. In this case $n_j=n$ for all $j$ in
(\ref{cs}) and the $B$-field acts as a source for only the untwisted 
R-R 2-form. 
As we have seen in (\ref{eigenM}), the associated field theory sector behaves 
as supersymmetric and hence stable. This suggest that perhaps in 
this case the field theory has an stable vacuum. 
It is natural then to make the following conjecture: whenever the bulk
tachyons condense and we end up with type IIB string theory on flat
background the 'twisted' sector tachyons of the field theory will
condense as well and we will end generically with $U(n)$ ${\cal N}=4$ SYM.
In that case the role of field theory 
the instability would be to avoid that the regular brane 
charge could decompose in its constituent physical branes.
It could be extremely interesting and useful if, by using such a conjecture 
and by proving that the field-theory tachyons condense as proposed, we 
could improve our understanding of closed string tachyon condensation.

\Acknowledgements

We would like to thank O. Bergman, J. Barb\'on, C. G\'omez, P. Horava, 
K. Landsteiner, L. Susskind, S. Theisen and P. Vanhove for useful discussions.
A. M. U. thanks M.~Gonz\'alez for kind encouragement and support.


\begin{thebibliography}{99}
\bibitem{Adams:2001sv}
A.~Adams, J.~Polchinski and E.~Silverstein,
``Don't panic! Closed string tachyons in ALE space-times,''
JHEP {\bf 0110}, 029 (2001)
[hep-th/0108075].

\bibitem{David:2001vm}
J.~R.~David, M.~Gutperle, M.~Headrick and S.~Minwalla,
``Closed string tachyon condensation on twisted circles,''
JHEP {\bf 0202}, 041 (2002)
[hep-th/0111212].

\bibitem{Douglas:2001ba}
M.~R.~Douglas and N.~A.~Nekrasov,
``Noncommutative field theory,''
Rev.\ Mod.\ Phys.\  {\bf 73}, 977 (2001)
[hep-th/0106048].

\bibitem{Szabo:2001kg}
R.~J.~Szabo,
``Quantum field theory on noncommutative spaces,''
hep-th/0109162.

\bibitem{Minwalla:2000px}
S.~Minwalla, M.~Van Raamsdonk and N.~Seiberg,
``Noncommutative perturbative dynamics,''
JHEP {\bf 0002}, 020 (2000)
[hep-th/9912072].

\bibitem{Matusis:2000jf}
A.~Matusis, L.~Susskind and N.~Toumbas,
``The IR/UV connection in the non-commutative gauge theories,''
JHEP {\bf 0012}, 002 (2000)
[hep-th/0002075].

\bibitem{Terashima:2000xq}
S.~Terashima,
``A note on superfields and noncommutative geometry,''
Phys.\ Lett.\ B {\bf 482}, 276 (2000)
[arXiv:hep-th/0002119].

\bibitem{Matsubara:2000gr}
K.~Matsubara,
``Restrictions on gauge groups in noncommutative gauge theory,''
Phys.\ Lett.\ B {\bf 482}, 417 (2000)
[hep-th/0003294].

\bibitem{Armoni:2000xr}
A.~Armoni,
``Comments on perturbative dynamics of non-commutative Yang-Mills theory,''
Nucl.\ Phys.\ B {\bf 593}, 229 (2001)
[hep-th/0005208].

\bibitem{Hayakawa:1999yt}
M.~Hayakawa,
``Perturbative analysis on infrared aspects of noncommutative QED on  R**4,''
Phys.\ Lett.\ B {\bf 478}, 394 (2000)
[hep-th/9912094].

\bibitem{Ruiz:2001hu}
F.~R.~Ruiz,
``Gauge-fixing independence of IR divergences in non-commutative U(1),  
perturbative tachyonic instabilities and supersymmetry,''
Phys.\ Lett.\ B {\bf 502}, 274 (2001)
[hep-th/0012171].

\bibitem{Landsteiner:2001ky}
K.~Landsteiner, E.~Lopez and M.~H.~Tytgat,
``Instability of non-commutative SYM theories at finite temperature,''
JHEP {\bf 0106}, 055 (2001)
[hep-th/0104133].

\bibitem{Guralnik:2002ru}
Z.~Guralnik, R.~C.~Helling, K.~Landsteiner and E.~Lopez,
``Perturbative instabilities on the non-commutative torus, Morita duality 
and twisted boundary conditions,''
JHEP {\bf 0205} (2002) 025
[hep-th/0204037].

\bibitem{Gomis:2000zz}
J.~Gomis and T.~Mehen,
``Space-time noncommutative field theories and unitarity,''
Nucl.\ Phys.\ B {\bf 591} (2000) 265
[hep-th/0005129].

\bibitem{Armoni:2001uw}
A.~Armoni and E.~Lopez,
``UV/IR mixing via closed strings and tachyonic instabilities,''
Nucl.\ Phys.\ B {\bf 632}, 240 (2002)
[hep-th/0110113].

\bibitem{Ibanez:1998qp}
L.~E.~Ibanez, R.~Rabadan and A.~M.~Uranga,
``Anomalous U(1)'s in type I and type IIB D = 4, N = 1 string vacua,''
Nucl.\ Phys.\ B {\bf 542}, 112 (1999)
[arXiv:hep-th/9808139].

\bibitem{Armoni:2002fh}
A.~Armoni, E.~Lopez and S.~Theisen,
``Nonplanar anomalies in noncommutative theories and the Green-Schwarz  mechanism,''
JHEP {\bf 0206}, 050 (2002)
[arXiv:hep-th/0203165].

\bibitem{Leigh:1998hj}
R.~G.~Leigh and M.~Rozali,
``Brane boxes, anomalies, bending and tadpoles,''
Phys.\ Rev.\ D {\bf 59} (1999) 026004
[hep-th/9807082].

\bibitem{VanRaamsdonk:2001jd}
M.~Van Raamsdonk,
``The meaning of infrared singularities in noncommutative gauge theories,''
JHEP {\bf 0111}, 006 (2001)
[hep-th/0110093].

\bibitem{Kiem:2001dm}
Y.~j.~Kiem, Y.~j.~Kim, C.~Ryou and H.~T.~Sato,
``One-loop noncommutative U(1) gauge theory from bosonic worldline  
approach,''
Nucl.\ Phys.\ B {\bf 630} (2002) 55
[hep-th/0112176].

\bibitem{Ishibashi:2000hs}
N.~Ishibashi, S.~Iso, H.~Kawai and Y.~Kitazawa,
``Wilson loops in noncommutative Yang-Mills,''
Nucl.\ Phys.\ B {\bf 573}, 573 (2000)
[hep-th/9910004].

\bibitem{Gross:2000ba}
D.~J.~Gross, A.~Hashimoto and N.~Itzhaki,
``Observables of non-commutative gauge theories,''
hep-th/0008075.

\bibitem{Das:2001ur}
S.~R.~Das and S.~P.~Trivedi,
``Supergravity couplings to noncommutative branes, open Wilson lines and  
generalized star products,''
JHEP {\bf 0102}, 046 (2001)
[hep-th/0011131].

\bibitem{Okawa:2001sh}
Y.~Okawa and H.~Ooguri,
``How noncommutative gauge theories couple to gravity,''
Nucl.\ Phys.\ B {\bf 599}, 55 (2001)
[hep-th/0012218].

\bibitem{Liu:2001ps}
H.~Liu and J.~Michelson,
``Supergravity couplings of noncommutative D-branes,''
hep-th/0101016.

\bibitem{Douglas:1996sw}
M.~R.~Douglas and G.~W.~Moore,
``D-branes, Quivers, and ALE Instantons,''
hep-th/9603167.

\bibitem{Seiberg:1999vs}
N.~Seiberg and E.~Witten,
``String theory and noncommutative geometry,''
JHEP {\bf 9909}, 032 (1999)
[hep-th/9908142].

\bibitem{Arcioni:2000bz}
G.~Arcioni, J.~L.~Barbon, J.~Gomis and M.~A.~Vazquez-Mozo,
``On the stringy nature of winding modes in noncommutative 
thermal field  theories,''
JHEP {\bf 0006}, 038 (2000)
[hep-th/0004080].

\bibitem{Murakami:2002yd}
K.~Murakami and T.~Nakatsu,
``Open Wilson lines as states of closed string,''
hep-th/0211232.

\bibitem{Kutasov:1990sv}
D.~Kutasov and N.~Seiberg,
``Number Of Degrees Of Freedom, Density Of States And Tachyons 
In String Theory And Cft,''
Nucl.\ Phys.\ B {\bf 358}, 600 (1991).

\bibitem{Doplicher:tu}
S.~Doplicher, K.~Fredenhagen and J.~E.~Roberts,
``The Quantum Structure Of Space-Time At The Planck Scale and Quantum 
Fields,''
Commun.\ Math.\ Phys.\ {\bf 172} (1995) 187.

\bibitem{Sen:1998ex}
A.~Sen,
``BPS D-branes on non-supersymmetric cycles,''
JHEP {\bf 9812} (1998) 021
[hep-th/9812031].

\bibitem{Jurco:2000dx}
B.~Jurco, P.~Schupp and J.~Wess,
``Nonabelian noncommutative gauge fields and Seiberg-Witten map,''
Mod.\ Phys.\ Lett.\ A {\bf 16} (2001) 343
[hep-th/0012225].


\bibitem{Zarembo:1999hn}
K.~Zarembo,
``Coleman-Weinberg mechanism and interaction of D3-branes in type 0
string theory,''
Phys.\ Lett.\ B {\bf 462}, (1999) 70
[hep-th/9901106].

\bibitem{Tseytlin:1999ii}
A.~A.~Tseytlin and K.~Zarembo,
``Effective potential in non-supersymmetric SU(N) x SU(N) gauge theory
and interactions of type 0 D3-branes,''
Phys.\ Lett.\ B {\bf 457}, (1999) 77
[hep-th/9902095].

\bibitem{Adams:2001jb}
A.~Adams and E.~Silverstein,
``Closed string tachyons, AdS/CFT, and large N QCD''
Phys.\ Rev.\ D {\bf 64}, (2001) 086001
[hep-th/0103220].

\end{thebibliography}
\end{document}